\documentclass[3p,times,twocolumn]{elsarticle}

\usepackage{ecrc}

\volume{00}

\firstpage{1}

\journalname{Nuclear Physics B Proceedings Supplement}

\runauth{}

\biboptions{numbers,sort&compress}

\jid{nuphbp}

\jnltitlelogo{Nuclear Physics B Proceedings Supplement}

\usepackage{amssymb,amsmath}

\usepackage{amssymb}

\usepackage[figuresright]{rotating}

\begin{document}

\begin{frontmatter}

\title{Heavy quarks and tau leptons: New physics opportunities}


\author{Alejandro Celis}

\address{IFIC, Universitat de Val\`encia -- CSIC, Apt. Correus 22085, E-46071 Val\`encia, Spain}

\begin{abstract}
In this talk I discuss the role of heavy quarks in new physics searches with tau leptons.   I focus on new physics effects associated to the scalar sector which are naturally enhanced for the heaviest fermions due to the large hierarchy of the fermion masses.    I will discuss two topics within this context: lepton flavour violation in the $\tau - \ell$ ($\ell=e,\mu$) sector and violations of lepton universality in tauonic $B$ decays.  
\end{abstract}


\end{frontmatter}


\section{Introduction}
\label{sec:intro}
Heavy quarks $(c,b,t)$ play an important role in many searches for physics beyond the Standard Model (SM) with tau ($\tau$) leptons.   I will focus here on the possibility of new physics effects associated to the scalar sector for which a connection between heavy quarks and the tau lepton is established due to their large mass.        The discovery of a SM-like Higgs boson by the ATLAS and CMS collaborations seems to confirm that a scalar sector is responsible for the breaking of the EW gauge symmetry~\cite{Aad:2012tfa,Chatrchyan:2012ufa}.       The strong hierarchy of the fermion mass spectrum implies that the SM Higgs couplings to light quarks and leptons are very suppressed.  The heaviest fermions offer then a unique place where exotic effects associated to the Higgs sector can be sizable due to their larger masses.         I will discuss two topics within this context:

\begin{itemize}

\item Lepton flavour violating (LFV) Higgs decays and Higgs mediated LFV $\tau$ decays.

\item Violation of lepton universality in $B\rightarrow D^{(*)} \tau \nu$ decays due to charged scalar contributions at tree level.

\end{itemize}

\section{Lepton flavour violation}
Let us assume that the recently discovered Higgs boson around $125$~GeV has lepton flavour violating couplings.  While such couplings are strongly constrained in the case of light leptons due to $\mu \rightarrow e \gamma$ and $\mu$ to $e$ conversion in nuclei searches, couplings of the type $\tau-\mu$ and $\tau-e$ receive much weaker constraints from the current limits on LFV $\tau$ decays~\cite{Blankenburg:2012ex,Harnik:2012pb}.     In Refs.~\cite{Harnik:2012pb,Davidson:2012ds} it was estimated that the LHC should be able to set stronger bounds on the hypothetical LFV Higgs couplings with the amount of data accumulated at the moment.    The interactions we are interested are described by
\begin{align} \label{lagGENN}
\mathcal{L}_{Y} \supset  - h \,\left\{  \, Y^{h}_{\tau \mu}  \left( \bar \tau \,  \mathcal P_R   \mu  \right)  +  Y^{h}_{\mu \tau}   \left( \bar \mu \,  \mathcal P_R  \tau  \right) + \mathrm{h.c.} \right\} \,.
\end{align}
The chirality projectors $ \mathcal P_{R,L} = (1 \pm \gamma_5)/2 $ are denoted as usual, the parameters $Y_{\tau \mu, \mu \tau}^{h}$ are taken as free complex quantities to be bounded experimentally.  Here we consider only $\tau-\mu$ LFV couplings but most of the arguments presented will hold also for  $\tau-e$.

The CMS collaboration has recently performed a search for LFV decays of the $125$~GeV Higgs boson~\cite{CMS:2014hha,Jan}, setting the upper bound on the branching ratio (BR) $\mathrm{BR}(h \rightarrow \tau \mu) \leq 1.57$ at 95\%~\text{CL.}\footnote{The CMS results present a slight excess of signal events with a significance of $2.5~\sigma$, the possibility to explain this excess within the framework of the type III 2HDM has been discussed in Ref.~\cite{Sierra:2014nqa}.}  The partial decay width for this process is given by
\begin{align}
\Gamma(h \rightarrow \tau \mu) \equiv& \;   \Gamma(h \rightarrow \tau^+ \mu^- + \mu^+ \tau^- ) \nonumber \\  =& \; \dfrac{ M_h }{  8 \pi}   \left(     |  Y_{\mu \tau}^{h} |^2 + |   Y_{ \tau \mu}^{h}|^2 \right)   \,.
\end{align}
Assuming that the total decay width of the Higgs boson is simply the SM Higgs total width plus $\Gamma(h \rightarrow \tau \mu)$ one can extract the limit  $\sqrt{   |Y_{\mu \tau}^{h}|^2  + |Y_{\tau \mu}^{h}|^2   } \leq 0.0036$~\cite{Jan}.      The top quark plays a very important role in the search for LFV Higgs decays.  Gluon fusion, together with vector boson fusion and associated Higgs production with a vector boson, constitute the relevant Higgs production mechanisms in the search for $h \rightarrow \tau \mu$ decays.   The SM Higgs coupling to gluons is described in the heavy top mass limit by the effective Lagrangian
\begin{equation} \label{nond}
\mathcal{L}_{ggh}\;=\;    \frac{\alpha_s}{12 \pi  v   }  G^{\mu \nu } G_{\mu \nu}   h \,.
  \end{equation}
Note that the dimension five operator $ G^{\mu \nu}  G_{\mu \nu} h $ comes suppressed by the EW scale $v = (\sqrt{2} G_F)^{-1/2} \simeq 246$~GeV instead of the top mass, showing a non-decoupling behavior.

Assuming that all the diagonal Higgs couplings are equal to their SM value one can translate the CMS limits on $\mathrm{BR}(h \rightarrow \tau \mu)$ into the following upper bounds:
\begin{align} \label{eq2n}
\mathrm{BR}( \tau \rightarrow \mu \gamma ) &\leq 2.3 \times 10^{-9} \,,\nonumber \\
\mathrm{BR}( \tau \rightarrow 3 \mu ) &\leq 4.7 \times 10^{-12} \,, \nonumber \\
\mathrm{BR}( \tau \rightarrow \mu \pi^+ \pi^- ) &\leq 1.5 \times 10^{-11} \,.
\end{align}
Within the scenario considered, the current CMS limit on $\mathrm{BR}(h\rightarrow \tau \mu)$ puts the three-body LFV $\tau$ decays beyond the reach of Belle II.

\begin{figure}[tb]
\centering
\includegraphics[width=3.3cm,height=1.7cm]{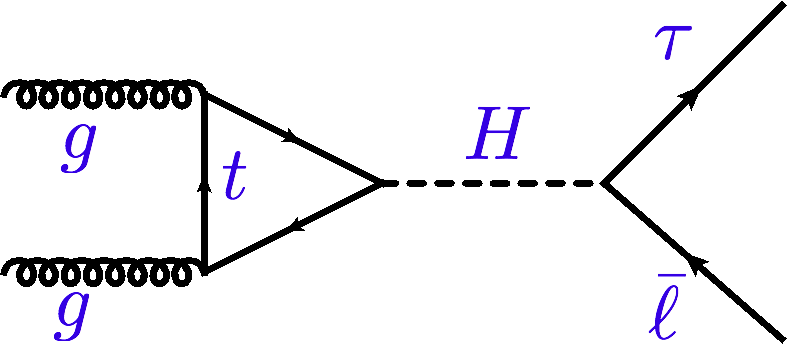}
~~ \qquad
\includegraphics[width=3.3cm,height=2.4cm]{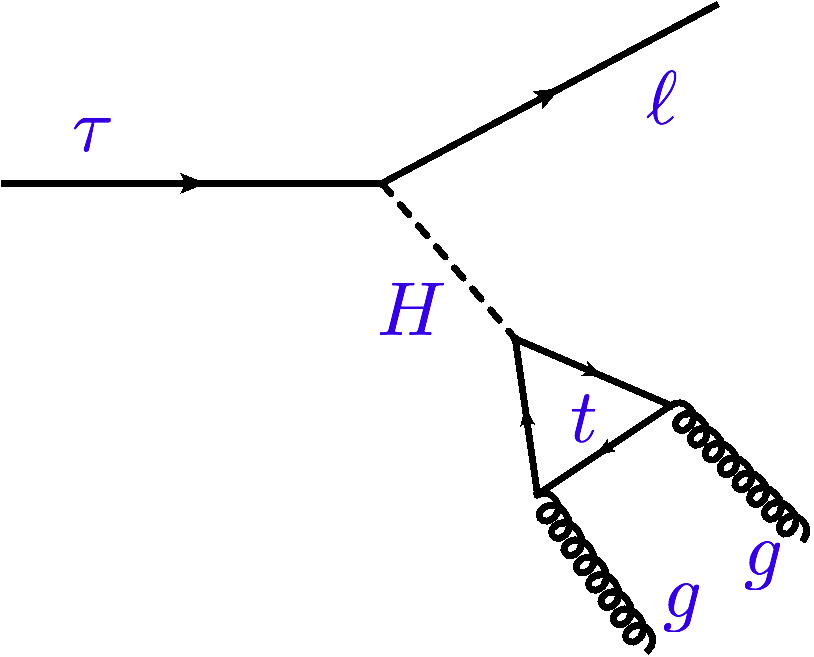}
(a) \qquad \qquad \qquad  \qquad  \qquad  ~~  (b)
\caption{\label{fig:introII} \it \small (a): Higgs production via gluon fusion and later decay into a pair $\tau-\ell$ with $\ell= e, \mu$.    (b)  Contribution to the LFV decay $\tau \rightarrow \ell \pi^+ \pi^-$ due to LFV scalar interactions. }
\end{figure}

Heavy quarks are crucial for the estimation of the $\tau \rightarrow \mu \gamma$, $\tau \rightarrow  3 \mu$ and $\tau \rightarrow  \mu \pi^+ \pi^-$ decay rates considered in Eq.~\eqref{eq2n}.     The same non-decoupling behavior of heavy quarks in Eq.~\eqref{nond} is also relevant at very low energies when describing Higgs mediated $\mu$ to $e$ conversion in nuclei~\cite{Shifman:1978zn,Crivellin:2014cta} and LFV  semileptonic $\tau$ decays~\cite{Celis:2013xja}, see Fig.~\ref{fig:introII}.

\begin{figure}[h]
\centering
\includegraphics[width=3.3cm,height=2.1cm]{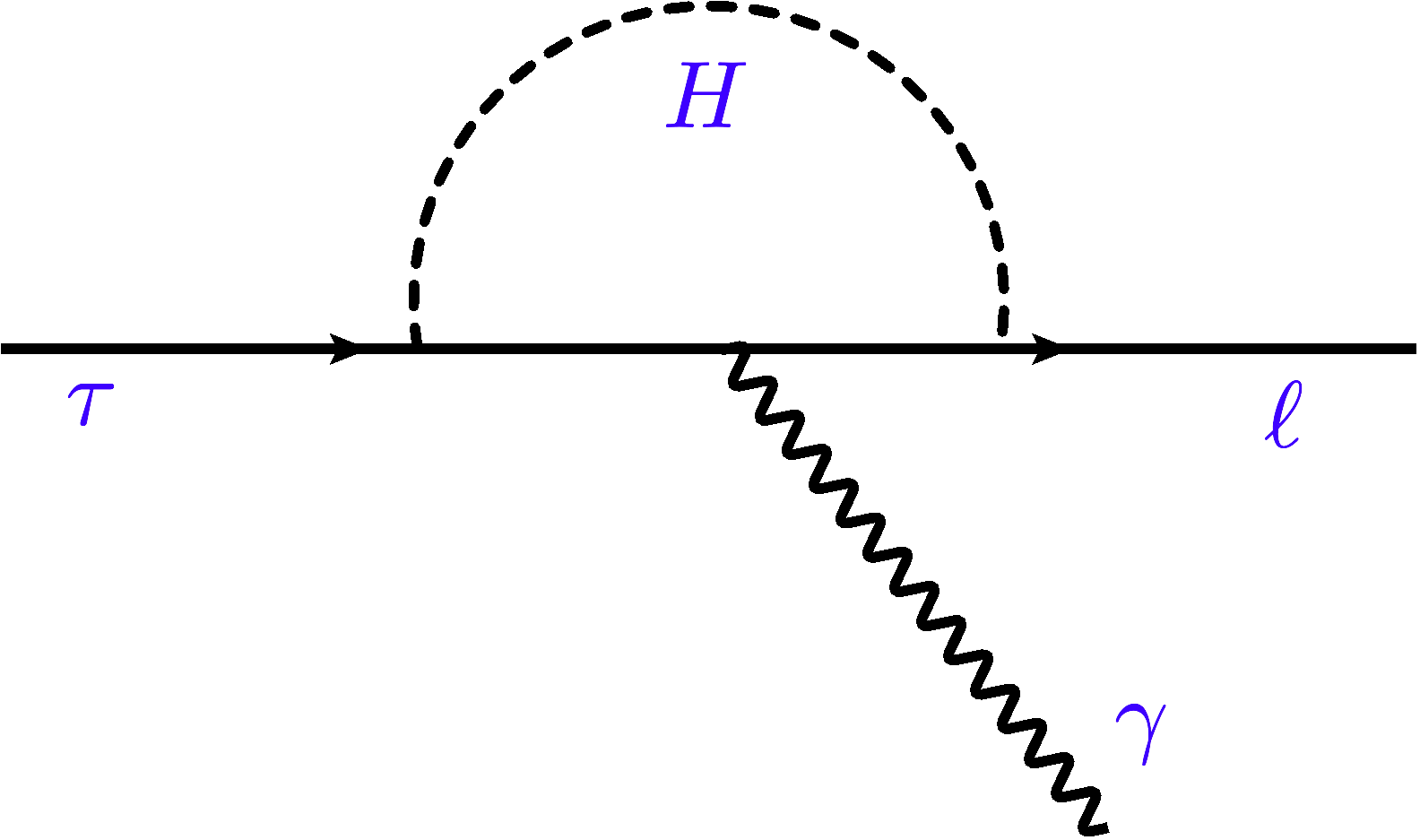}
~
\includegraphics[width=3.3cm,height=2.3cm]{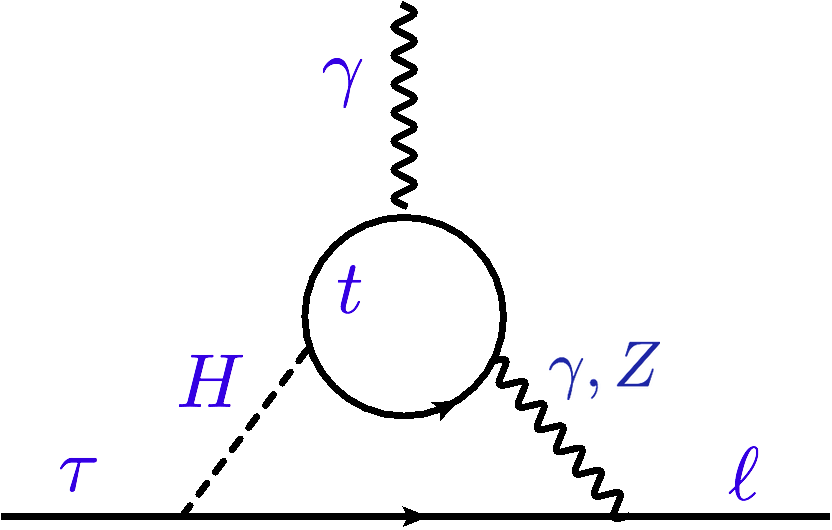} \\
(a) \qquad \qquad \qquad    \qquad  ~~  (b)
\caption{\label{fig:intro} \it \small (a) Scalar mediated one-loop contribution to $\tau \rightarrow \ell \gamma$.  (b) Two-loop contribution of the Barr-Zee type to $\tau \rightarrow \ell \gamma$.}
\end{figure}

\begin{figure}[h]
\centering
\includegraphics[width=7.9cm,height=5.9cm]{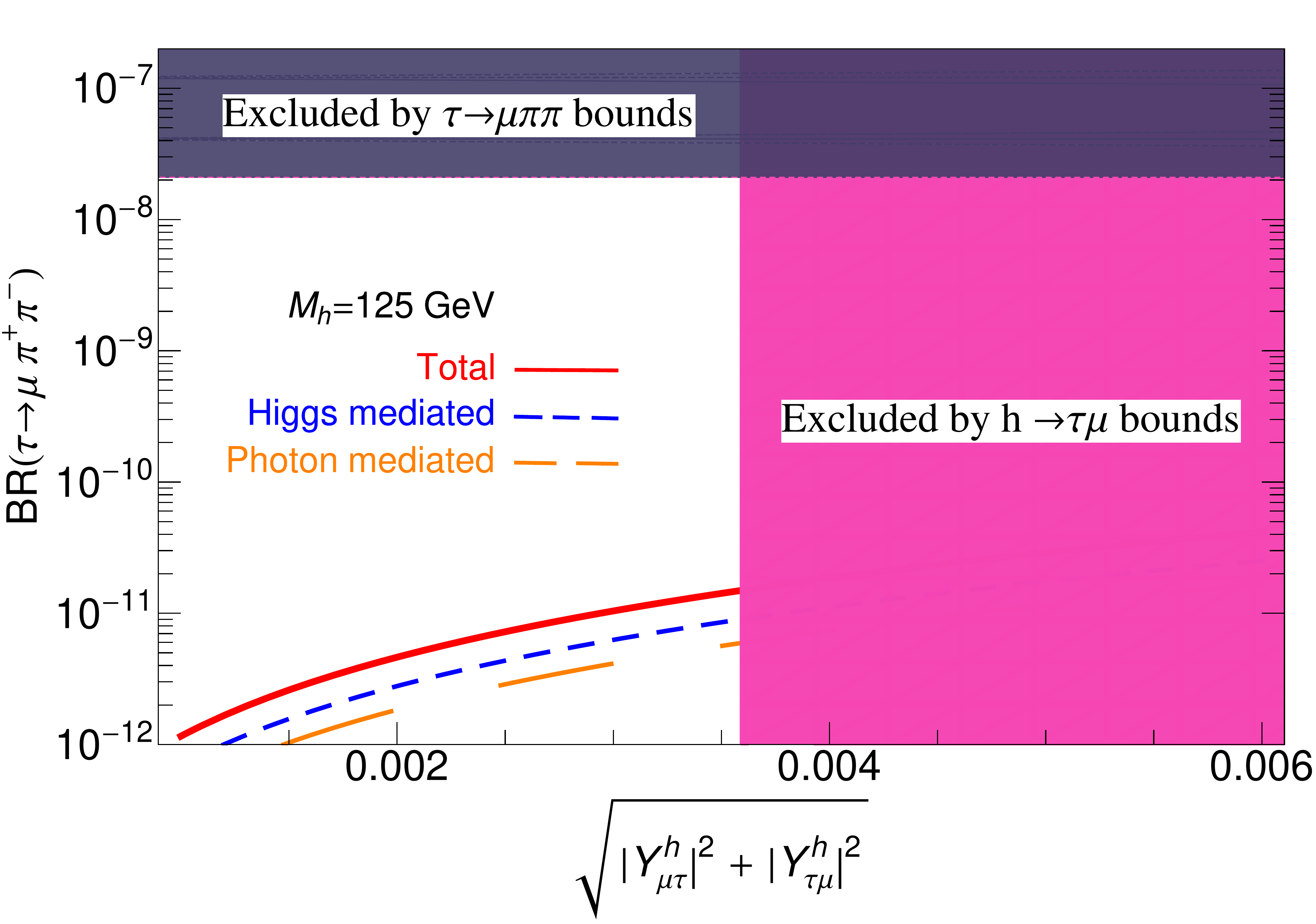}
\caption{\label{fig:gluonic} \it \small Constraints on LFV Higgs couplings from $\tau \rightarrow \mu \pi^+ \pi^-$ searches at B-factories and LFV Higgs decays at the LHC.    We assume that the $125$~GeV Higgs has LFV couplings and take all diagonal Higgs couplings to be SM-like as a benchmark.   Figure adapted from Ref.~\cite{Celis:2013xja}.     }
\end{figure}

The LFV radiative decay $\ell \rightarrow \ell^{\prime} \gamma$ is described by the effective Lagrangian, 
\begin{equation} \label{eq:dipole}
\mathcal{L}_{\mbox{\scriptsize{eff}}} = -  \frac{m_{\tau}}{\Lambda^2}     \left\{   \left( \mathrm{C_{DR}}  \bar \mu \sigma^{\rho \nu}  \mathcal{P}_L \tau    + \mathrm{C_{DL}}  \bar \mu \sigma^{\rho \nu}  \mathcal{P}_R \tau \right) F_{\rho \nu} + \mathrm{h.c.}    \right\} .
\end{equation}
Here $\Lambda$ represents the high energy scale where the new physics appears.    The dimension five operators in Eq.~\eqref{eq:dipole} appear at the loop level in renormalizable theories.   Dipole transitions require a chirality flip between the initial fermion state and the final one.   Scalar mediated contributions to $\ell \rightarrow \ell^{\prime} \gamma$ at the one-loop level involve three chirality flips, two in the Yukawa couplings and one in the fermion propagator, see Fig.~\ref{fig:intro} (a).   This accidental suppression is avoided at the two-loop level in diagrams of the Barr-Zee type~\cite{Bjorken:1977vt,Barr:1990vd,Chang:1993kw}.     In particular, the Wilson coefficients $\mathrm{C_{DL (DR)}}$ receive large contributions from two-loop diagrams with intermediate heavy quarks as the one shown in Fig.~\ref{fig:intro} (b).    Similar contributions also enter in the three-body decays $\tau \rightarrow 3 \mu$ and $\tau \rightarrow \mu \pi^+ \pi^-$.  The complementarity between low energy searches for LFV via $\tau \rightarrow \mu \pi^+ \pi^-$ decays and searches for LFV Higgs decays is summarized in Fig.~\ref{fig:gluonic}.

\section{Lepton universality}
The BaBar collaboration has reported an excess with respect to the Standard Model (SM) in exclusive semileptonic transitions of the type $b \rightarrow c \tau^- \bar \nu_{\tau}$~\cite{Lees:2012xj}.  More specifically, they have measured the ratios
\begin{align}  
R(D) \equiv \dfrac{   \mathrm{BR}(   \bar B \rightarrow D \tau^- \bar \nu_{\tau} ) }{   \mathrm{BR(  \bar B \rightarrow D \ell^-  \bar \nu_{\ell} )} }  =\;  0.440 \pm 0.058 \pm 0.042 \,, \nonumber \\
R(D^*) \equiv \dfrac{ \mathrm{BR}(\bar B \rightarrow D^* \tau^- \bar \nu_{\tau})  }{  \mathrm{BR}(\bar B \rightarrow D^* \ell^- \bar \nu_{\ell}) }  =\; 0.332 \pm 0.024 \pm 0.018 \,,
\end{align}
normalized by the corresponding light lepton modes~$\ell = e, \mu$.     These measurements present an excess of $2.0 \sigma$ ($R(D)$) and $2.7 \sigma$ ($R(D^*)$) with respect to the SM~\cite{Lees:2012xj,Kamenik:2008tj,Fajfer:2012vx}.        This excess would imply a violation of lepton universality at the $\sim30\%$ level.  The most natural explanation to this excess would be a new charged mediator entering at tree-level which couples preferentially to the tau lepton.

\begin{figure}[tb]
\centering
\includegraphics[width=4.4cm,height=3.4cm]{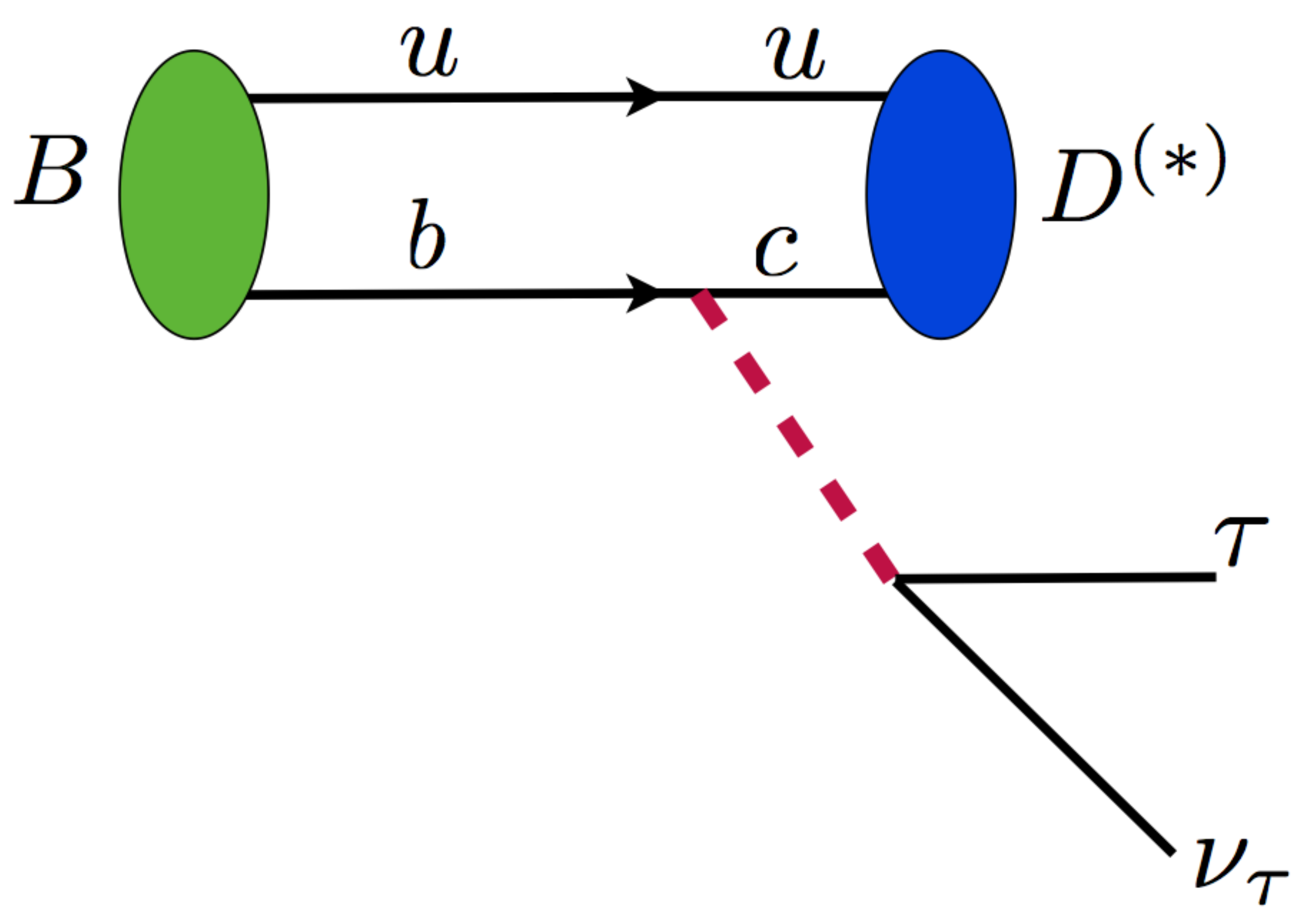}
\caption{\label{fig:semi} \it \small Charged scalar contribution to the exclusive semileptonic decay $B \rightarrow D^{(*)} \tau \nu$ (pink-dashed propagator).}
\end{figure}

Multi-Higgs doublet models predict the existence of charged scalars which would contribute at tree level in $B \rightarrow D^{(*)} \tau \nu$ decays, see Fig.~\eqref{fig:semi}.     Moreover, by imposing the natural flavour conservation condition (NFC) to avoid dangerous flavour changing neutral currents one obtains that the charged scalar couplings to fermions are proportional to the fermion masses.     I will consider here the aligned two-Higgs-doublet model (A2HDM)~\cite{Pich:2009sp}; charged Higgs couplings are still proportional to the fermion masses in this framework and all the versions of the 2HDM with NFC are recovered in particular limits.    Charged Higgs interactions with fermions are parametrized in the A2HDM by
\begin{align}\label{lagrangian}
 \mathcal{L}_Y  =&  - \dfrac{\sqrt{2}}{v}\; H^+ \Bigl\{ \bar{u} \left[ \varsigma_d\, V_{\mbox{\scriptsize{CKM}}} M_d \mathcal P_R - \varsigma_u\, M_u V_{\mbox{\scriptsize{CKM}}} \mathcal P_L \right]  d\,  \nonumber \\
 &+ \, \varsigma_l\, \bar{\nu} M_l \mathcal P_R l \Bigr\}
\;  + \;\mathrm{h.c.} \, .
\end{align}
The parameters $\varsigma_{f}$ ($f=u,d,l$) are complex family universal parameters that need to be determined by experimental data.     The matrix $V_{\mbox{\scriptsize{CKM}}}$ represent the quark mixing matrix and $M_{f}$ ($f=u,d,l$) are the diagonal fermion mass matrices.   
\begin{figure}[tb]
\centering
\includegraphics[width=7.3cm,height=6.3cm]{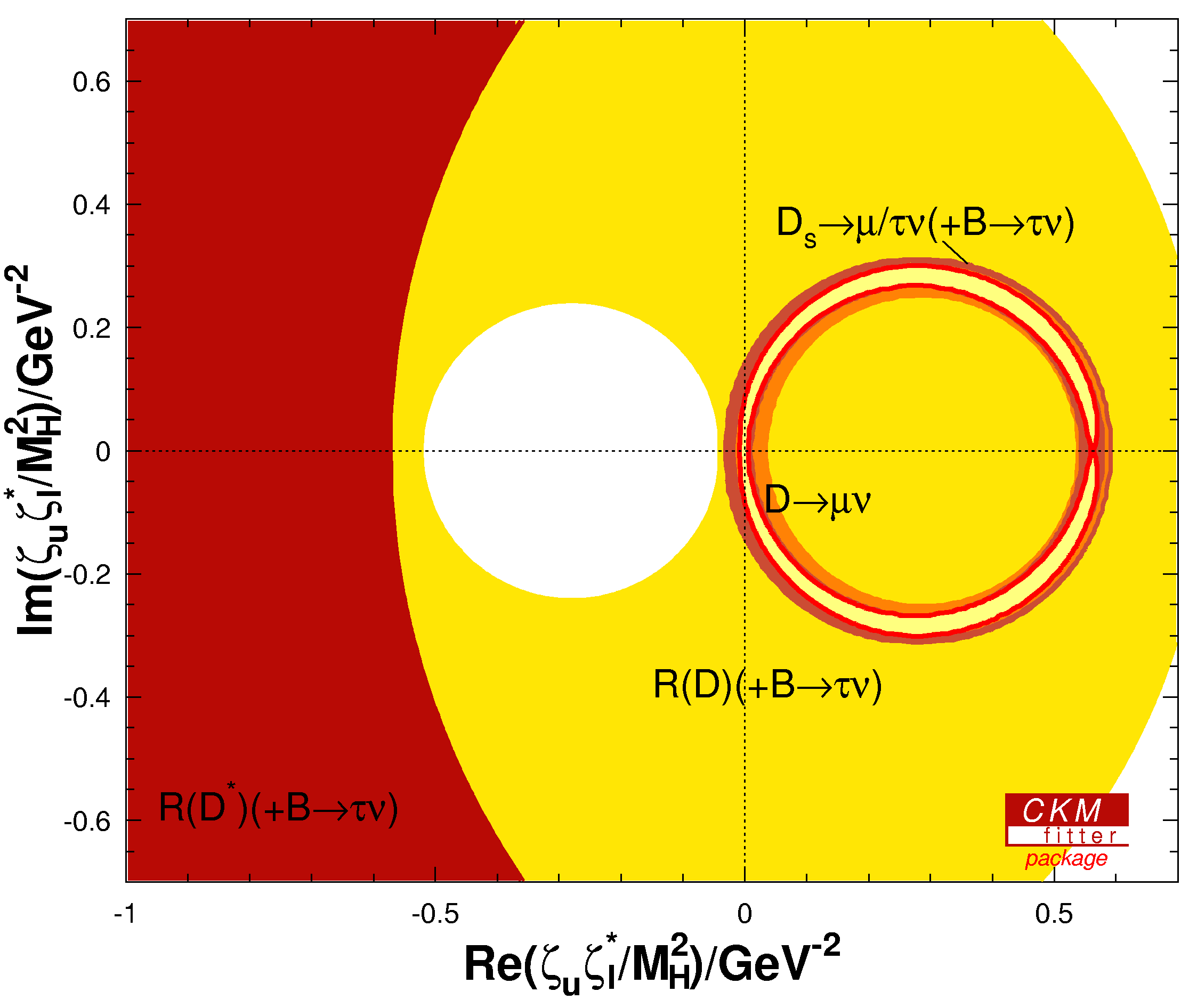}
\caption{\label{fig:tr} \it \small
$95\%$~CL allowed regions in the plane $\varsigma_u \varsigma_l^*/M_{H^{\pm}}^2$ considering different observables which receive a charged Higgs contribution at tree level.}
\end{figure}

Considering leptonic $B$ and $D_{(s)}$ meson decays in which the charged Higgs also enters at tree level, it is not possible to accommodate the current excess in $R(D^{(*)})$ within the A2HDM~\cite{Celis:2012dk}.    Fig.~\ref{fig:tr} shows the allowed regions by the different observables considered in the $\varsigma_u \varsigma_l^*/M_{H^{\pm}}^2$ plane.  It can be observed that the preferred values by $R(D^*)+B \rightarrow \tau \nu$ are in conflict with the allowed region by $D_{(s)}$-meson leptonic decays.   As a by-product of this result, we know that all versions of the 2HDM with NFC (this includes the type II 2HDM) cannot accommodate the excess in $R(D^{(*)})$.       Interestingly, all the observables can be accommodated within the A2HDM if $R(D^*)$ is not considered in the fit.    The reason for this is that $R(D)$ is sensitive to the scalar combination of charged Higgs couplings while leptonic $B$ and $D_{(s)}$ decays are sensitive to the pseudoscalar combination.    Due to the vectorial nature of the $D^*$, $R(D^*)$ is sensitive to the pseudoscalar combination which is already constrained by leptonic meson decays.

If the current excess in exclusive semileptonic $b \rightarrow c \tau^- \bar \nu_{\tau}$ transitions persists, it will be crucial to measure differential distributions in these decays.    Differential distributions in the momentum transfer and polarization fractions of the $D^*$ meson can be very useful to disentangle different kinds of new physics, see Refs.~\cite{Kamenik:2008tj,Fajfer:2012vx,Celis:2012dk,Datta:2012qk,Biancofiore:2013ki} and references therein.

\section{Conclusions}
Heavy quarks $(c,b,t)$ play a crucial role in many circumstances when searching for new physics using $\tau$ leptons.    I have focused on new physics effects associated to the scalar sector.    Due to their large masses compared with that of the other fermions, heavy quarks and the $\tau$ lepton are expected to have stronger couplings with scalar particles from an extended Higgs sector.

I discussed the role of heavy quarks when probing for $\tau-\ell$ ($\ell=e, \mu$) lepton flavour violation associated to the scalar sector via Higgs or $\tau$ decays.      The non-decoupling behaviour of the top-quark in the effective $h gg$ vertex is crucial for the search of $h \rightarrow \tau \ell $ ($\ell = e, \mu$) at the LHC.     Heavy quarks play also a special role in the description of LFV $\tau$ decays because they enter in two-loop diagrams of the Barr-Zee type as the one shown in Fig.~\ref{fig:intro} (b).    The one-loop scalar mediated contribution shown in Fig.~\ref{fig:intro} (a) is accidentally suppressed so that two-loop diagrams can dominate over the one-loop contribution.

I have also discussed possible violations of lepton universality in $B \rightarrow D^{(*)} \tau \nu$ decays, taking into account the observed excess by the BaBar collaboration in these modes.     The current excess in $R(D^{*})$ cannot be accommodated within the framework of the A2HDM~\cite{Celis:2012dk}.  As a by-product of this analysis one obtains that none of the two-Higgs-doublet models with natural flavour conservation (this includes the type II 2HDM) can explain this excess either.   Recent theoretical developments have reduced the discrepancy between theory and experiment in $R(D)$ to about $1\sigma$~\cite{Bailey:2012jg,Becirevic:2012jf}.      The situation for $R(D^*)$ is still unclear.    New Belle measurements of $R(D)$ and $R(D^*)$ will be useful to clarify the current situation.    Future measurements of differential distributions in $B \rightarrow D^{(*)} \tau \nu$ decays would provide a crucial handle to discriminate between different new physics alternatives in case the current excess in $R(D^{(*)})$ persists.

\end{document}